\documentclass[a4paper,nofootinbib,onecolumn]{revtex4-1}
\usepackage[utf8]{inputenc}
\usepackage{graphicx}
\usepackage{physics}
\usepackage{hyperref}
\usepackage{amsmath,amssymb}
\usepackage{bbold}
\usepackage{dsfont}

\begin{document}
\title{Quantum steering and quantum discord under noisy
channels and entanglement swapping}

\author{Pedro Rosario$^1$}
\email{prosariovargas@correo.unicordoba.edu.co}

\author{Andr\'es F. Ducuara$^{2,3,4}$}
\email{andres.ducuara@bristol.ac.uk}

\author{Cristian E. Susa$^{1,5}$}
\email{cristiansusa@correo.unicordoba.edu.co}

\address{$^1$Department of Physics and Electronics, University of C\'ordoba, 230002 Monter\'ia, Colombia}
 
\address{$^2$Quantum Engineering Centre for Doctoral Training, University of Bristol, Bristol BS8 1FD, United Kingdom}

\address{$^3$Quantum Engineering Technology Labs, University of Bristol, Bristol BS8 1FD, United Kingdom}
  
\address{$^4$Department of Electrical and Electronic Engineering, University of Bristol, Bristol BS8 1FD, United Kingdom}
 
\address{$^5$Centre for Bioinformatics and Photonics---CIBioFi, Calle 13 No.~100-00, Edificio 320 No.~1069, 760032 Cali, Colombia}


\begin{abstract}
Quantum entanglement, quantum discord, and EPR-steering are properties which are considered as valuable resources for fuelling quantum information-theoretic protocols. EPR-steering is a property that is more general than Bell-nonlocality and yet more restrictive than entanglement. Quantum discord on the other hand, captures non-classical behaviour beyond that of entanglement, and its study has remained of active research interest during the past two decades. Exploring the behaviour of these quantum properties in different physical scenarios, like those simulated by open quantum systems, is therefore of crucial importance for understanding their viability for quantum technologies. In this work, we analyse the behaviour of EPR-steering, entanglement, and quantum discord, for two-qubit states under various quantum processes. First, we consider the three \emph{noisy channel} scenarios of; phase damping, generalised amplitude damping and stochastic dephasing channel. Second, we explore the behaviour of these quantum properties in an \textit{entanglement swapping} scenario. We quantify EPR-steering by means of an inequality with three-input two-output measurement settings, and address quantum discord as the interferometric power of quantum states. Our findings are the following. First, we show that some of the relatively straightforward noisy channels here considered, can induce non-trivial dynamics such as sudden death as well as death and revival of EPR-steering and entanglement. Second, we find that although noisy channels in general reduce the amount of correlations present in the system, the swapping protocol on the other hand displays scenarios where these quantum correlations can be enhanced. These results therefore illustrate that quantum processes do not exclusively affect the quantum properties of physical systems in a negative manner, but that they can also have positive effects on such properties.

\end{abstract}
 
\maketitle



\section{Introduction}
\label{introduction}


Quantum features such as coherence and correlations play a fundamental role in quantum foundations as well as in practical matters like the development of quantum technologies  \cite{pramanik2019, wang2021, chitambar2019, garding2021}. In particular, quantum optics stands out as a key field of research, due to photonic devices allowing for the experimental implementation of quantum information-theoretic protocols, where quantum correlations can be detected, tested, and exploited \cite{Su2021, huang2021, zhao2020}. There exists nowadays a whole zoo of different quantum properties which are deemed as potential resources\footnote{
It is worth noting that, in addition to properties of states, properties of quantum objects like channels and measurements can also be considered as potential resources.
}, 
and which can be broadly classified as: i) \emph{entanglement-based} correlations, meaning correlations which necessarily require the presence of entanglement for them to exist and ii) \emph{above-entanglement} correlations, which aim at capturing non-classicality beyond that of entanglement. Entanglement-based correlations include properties like: a) Bell-nonlocality, which is arguably the most restrictive amongst these properties \cite{bell1964}, b) EPR-steering, which is conceived as the possibility of remotely generating ensembles of quantum states not obeying local hidden state (LHS) models \cite{wiseman2007}, and which has become of great research interest due to its usefulness for \textit{device-independent} protocols \cite{zhao2020, branciard2012},  and c) usefulness for teleportation, which provides information on how useful a state is for the standard teleportation scheme \cite{Horodecki_1996}. Above-entanglement correlations on the other hand consist on properties which are more general than entanglement. Quantum discord, as named by its discoverers \cite{zurek2001}, is one of such correlations \cite{ferraro2010}. Theoretical and experimental developments on quantum discord have remained of great research interest during the past two decades, with a plethora of different measures being proposed for its quantification \cite{adesso2016}. Another type of quantum property beyond entanglement which has recently been introduced is the quantum obesity, which relates to the volume of the steering ellipsoids of bipartite quantum states \cite{Milne_2014, Jevtic_2014}. 

These quantum properties have been explored in various different setups and physical systems, for instance, in the generation of X-form quantum states on quantum computers \cite{garding2021}. The exploration and analysis of the behaviour of these quantum properties, under different physical scenarios, continues to be of relevant research interest, this, mostly due to the ongoing efforts to better understand and harness these quantum properties, which has been reflected in a great diversity of techniques for quantification, detection and manipulation \cite{ducuara2020sudden, yadin2021}. Optical systems, for instance, have extensively been used as a platform to experimentally investigate quantum properties such as EPR-steering \cite{yang2021, bart2016} and quantum discord \cite{xu2017, allati2014}.

In this work we concentrate our efforts to explore three quantum properties: i) entanglement, ii) EPR-steering, as an entanglement-based correlation, and iii) quantum discord, as an above-entanglement correlation, in bipartite two-qubit states affected by various quantum processes. Specifically, we report on the dynamics of EPR-steering by means of the Costa-Angelo criterion \cite{costa2016}, quantum discord quantified by means of the interferometric power of quantum states \cite{girolami2014}, and entanglement quantified by the concurrence \cite{Wootters_1998}. We present a systematic analysis of these quantum features under different noisy channel scenarios, as well as under entanglement swapping protocols. First, we show that relatively straightforward noisy channels can still induce nontrivial dynamics in the form of sudden death as well as a death and revival of EPR-steering and entanglement. Second, we show that whilst noisy channels generally reduce the amount of quantum properties present in the system, swapping protocols on the other hand can increase the amount of such properties, even to the maximal amount allowed by quantum theory. These results therefore illustrate how quantum processes can affect the quantum properties of physical systems in both negative and positive manners.

This document is organised as follows: in Section \ref{qcorr}, we introduce measures for quantifying EPR-steering, entanglement and discord. In Section \ref{noise-swapping} we briefly describe the operator-sum-representation formalism, the set of noisy channels here considered, and the entanglement swapping protocol implemented to discuss the behaviour of these quantum properties. In Section \ref{results}, we present an analysis and discussion on the behaviour of these quantum properties under the different quantum processes, and focus on the amount of EPR-steering, discord, and entanglement that is possible to be transmitted over these processes. Finally, we derive conclusions in Section \ref{conclusion}.

\section{Quantum steering and quantum discord}
\label{qcorr}

Quantum states can be described by means of the density operator formalism. For the bipartite (two parties) case, we denote the density operator as $\rho\equiv\rho_{AB}$, living in a Hilbert space $\mathcal{H}^{d_A}\otimes\mathcal{H}^{d_B}$, where $d_{A(B)}$ is the dimension of the subsystem (party) $A(B)$, respectively. In this work we consider two-qubit states, meaning $d_A=d_B=2$. The three quantum properties as well as the set of quantum states of interest in this work are introduced as follows.

\subsection{Entanglement by means of the Concurrence}

A bipartite quantum system is said to be entangled if its state cannot be written as a statistical mixture of tensor product of local states of each composing party. One unambiguous quantifier of entanglement for general two-qubit states is the {\it entanglement of formation}, that is conceived as the minimum amount of pure entangled states required in the decomposition of the state of interest. However, an auxiliary quantity that is commonly used to quantify entanglement is the {\it concurrence}, which is a monotonic function of the entanglement of formation \cite{Wootters_1998}. For mixed states, concurrence is computed as:
\begin{equation}
    C=\max\left\{0,\sqrt{\lambda_1}-\sqrt{\lambda_2}-\sqrt{\lambda_3}-\sqrt{\lambda_4}\right\} ,
    \label{eq:concur}
\end{equation}
where $\lambda_1> \lambda_2 > \lambda_3 > \lambda_4$ are the eigenvalues of  $\rho(\sigma_{y}\otimes\sigma_{y})\rho^{*}(\sigma_{y}\otimes \sigma_{y})$, and $\rho^*$ is the complex conjugate of the density operator. $C \in [0,1]$ and, in particular, $C=0$ for separable states, while $C=1$ for Bell (maximally entangled) states.

\subsection{EPR-steering by means of the Costa-Angelo criterion}

EPR-steering is a quantum property less restrictive than Bell-nonlocality and yet more restrictive than entanglement. In the (2,2,2)-scenario, meaning two parties, two measurements, and two outcomes per measurement, it turns out that the EPR-steering inequality related to this scenario and its maximum violation for two-qubit systems leads to the same result that the well-known CHSH-inequality \cite{costa2016}. In the next particular scenario where the steering part has access to three measurements, an additional non-trivial steering inequality appears, which we address here as the $\rm F_3$-inequality \cite{cavalcanti2009}. In the two-qubit case, the Costa-Angelo criterion \cite{costa2016} establishes that a general two-qubit state violates the $\rm F_3$-inequality if and only if:
\begin{align}
    s_1^2+s_2^2+s_3^2>1,
    \label{eq:steering-inequality}
\end{align}
with $\{s_1, s_2, s_3\}$ the singular values of the correlation matrix $T=[T_{ij}]$, with entries $T_{ij}={\rm{Tr}}[\rho (\sigma_i  \otimes \sigma_j)]$, and $\sigma_i$, $i=1, 2, 3$ the Pauli matrices. As an indicator of how much steerable a quantum state is, we compute the quantifier \cite{costa2016}:
\begin{align}
    F_3(\rho)=\max\left\{0,\frac{\sqrt{s_1^2+s_2^2+s_3^2}-1}{\sqrt{3}-1}\right\}.
    \label{eq:f3}
\end{align}
Interestingly, EPR-steering has allowed for the development of a geometric technique to visualise states of two qubits by means of a concept called ``steering ellipsoid", which is constructed with all the possible states that Bob can ``steer" to Alice after carrying out local measurements on his party. Through the properties of this ellipsoid, like volume, orientation and shape, it is possible to geometrically interpret various aspects of entanglement and discord \cite{Jevtic_2014}.

\subsection{Discord as the interferometric power of quantum states}

Quantum discord is considered amongst the most general forms of quantum correlations known so far, and has received a lot of attention during the last two decades (see e.g., \cite{vedral2012, bera2017, zurek2001, girolami2011, davidovich2010}). Many quantifiers of discord have been proposed according to the bona fide criteria for quantum correlations \cite{adesso2016}. In this work, we make use of a definition of  discord
as the {\it interferometric power} of quantum states proposed in the context of quantum metrology {\cite{girolami2014}}. It is defined as
\begin{align}
   \text{IP}(\rho_{AB})=\min_{H_{A}} F(\rho_{AB};H_{A}),
    \label{eq:IP}
\end{align}
where $F$ is the quantum Fisher information (QFI) {\cite{Braunstein_1994}} and the minimum is taken over all local Hamiltonians with non-degenerate spectrum. For the case in which one system's party is a qubit, Eq. \eqref{eq:IP} reduces to
\begin{align}
   \text{IP}(\rho_{AB})=\zeta_{min}[M],
    \label{eq:IP_2}
\end{align}
 where $\zeta_{min}[M]$ is the smallest eigenvalue of the $3\times 3$ matrix $M$ with elements 
\begin{align}
 M_{m,n}=\frac{1}{2}\sum_{i,l; q_{i}+q_{l}\neq 0}\frac{(q_{i}-q_{l})^{2}}{q_{i}+q_{l}}\bra{\psi_{i}}\sigma_{mA}\otimes \mathbb{1}_{B}\ketbra{\psi_{l}}{\psi_{l}}\sigma_{nA}\otimes \mathbb{1}_{B} \ket{\psi_{i}},
    \label{eq:M}
\end{align}
with $\{q_{i},\ket{\psi_{i}}\}$ being the eigensystem of $\rho_{AB}$. In particular, Eq. \eqref{eq:M} vanishes if and only if $\rho_{AB}$ is either a classical-classical or a quantum-classical state. The operational interpretation of this discord is as follows: a zero-discorded state cannot guarantee a precision in parameter estimation in the worst case scenario, while discordant states are suitable for estimating parameters encoded by unitary transformations \cite{Braun_2018, girolami2014}.

\subsection{Quantum states of interest}

We consider a X-form structure for the density operator in order to discuss the behaviour of the aforementioned quantum properties:
\begin{eqnarray}
    \label{eq:xform}
    \rho_X=
    \begin{pmatrix} 
    a &0&0&w \\ 0 & b&z&0\\0&z^*&c&0 \\w^*&0&0&d
    \end{pmatrix} ,
\end{eqnarray}
where it is required that $a+b+c+d=1$. An interesting thing here is that Eq. \eqref{eq:xform} allows nice close formula for both concurrence; $C=2\max\left\{0,\abs{z}-\sqrt{ad},\abs{w}- \sqrt{bc}\right\}$ and EPR-steering; $F_3=\max\left\{0,\frac{\sqrt{(a - b - c + d)^2+8(|w|^2+|z|^2)}-1}{\sqrt{3}-1}\right\}$. 

In spite of its simplified form, Eq. \eqref{eq:xform} still represents an infinite set of quantum states. Hence, we pay particular attention to the following X-form quantum states, which we address here as Almeida et. al. states \cite{mafalda2007}, but which are also known as ``partially entangled states with coloured noise" \cite{LHV2018}, and are given by: 
\begin{equation}
    \rho(k,\theta)=k\ketbra{\phi^{+}(\theta)}{\phi^{+}(\theta)}+(1-k)\rho_{A}(\theta)\otimes (\mathds{1}/2),
    \label{eq:state}
\end{equation}
where $\ket{\phi^{+}(\theta)}=\cos{\theta}\ket{00}+\sin{\theta}\ket{11}$, 
$\rho_{A}(\theta)=\mathrm{Tr}_{B}\hspace{2pt}[\ketbra{\phi^{+}(\theta)}{\phi^{+}(\theta)}]$, and $0\leq \theta \leq \pi/4$. States in Eq. \eqref{eq:state} can be obtained by applying local filters on one party (party $B$ in this case) of a Werner state \cite{mafalda2007}. It is worth noting that local filters lead to enhancement of some quantum properties such as nonlocality and steering \cite{pramanik2019}. Almeida et. al. states are partially quantum correlated with probability $k$ (mixture parameter), while reduces to a classical state with probability $k-1$. In the particular case of $k=1$ and $\theta=\pi/4$, Almeida et. al. state reduces to the maximum correlated (Bell's) state $\ket{\phi^{+}(\pi/4)}=\frac{1}{\sqrt{2}}(\ket{00}+\ket{11})$. Another important aspect of the Almeida et. al. states Eq. \eqref{eq:state} is that it admits a local-hidden-state (LHS) model when the following condition is satisfied for $\theta$ and $k$ {\cite{Bowles_2016}}
\begin{align}
    \cos^{2}(2\theta)\geq \frac{2k-1}{(2-k)k^{3}}.
    \label{eq:uns1}
\end{align}
In other words, Almeida et. al. states are unsteerable when condition Eq. \eqref{eq:uns1} takes place.

\section{Quantum processes}
\label{noise-swapping}

In this section we present two kind of quantum processes which we consider to affect the set of quantum Almeida et. al. states in Eq. \eqref{eq:state}. At a first instance, we consider three noisy channels; phase damping, generalised amplitude damping and stochastic dephasing channel. All of them described in the operator-sum representation. On the second stage, we explore a quantum information transferring process, that is, the well-known entanglement swapping protocol. As in the latter process, it is needed four parties, we assume to have a couple of two-qubit systems in the same Almeida et. al. state.

\subsection{Noisy channels}

Operator-sum representation is used to describe the effect of the three considered quantum noises. Thus, a quantum state $\rho$ is transformed as \cite{kraus1983}:
\begin{equation}
   \rho'=\sum_{i}K_{i}\rho K_{i}^{\dagger} ,
\end{equation}
where Kraus operators $K_{i}$ satisfy the closure relation  $\sum_{i}K^{\dagger}_{i}K_{i}=I$, and in the context of open quantum systems, contain the information of the environmental effects on the system of interest.

\subsubsection{Phase damping (PD)} describes the loss of quantum information, for instance, photon scattering, without loss of energy. This is also known as phase flip channel. For a single qubit, Kraus operations can be written as {\cite{nielsen2000}}:
\begin{align}
K_1=\begin{pmatrix} 1 & 0 \\ 0 & \sqrt{1-p}
\end{pmatrix} \ \ \ \ \ K_2=\begin{pmatrix} 0 & 0 \\ 0 & \sqrt{p}
\end{pmatrix},
    \label{eq:Kraus_PD}
\end{align}
where $p\in[0,1]$ determines how coherence is loss as it increases to $1$. As our setup is composing of two qubits, we assume each of them to be affected equally by a PD noise; the corresponding four Kraus operators are constructed through tensor products as: $K_i\otimes K_j$, with $i,j=1,2$.

\subsubsection{Generalized amplitude damping (GAD)} describes dissipation of a system to its environment at a finite temperature, for example, relaxation of a spin due to its interaction with a bath or lattice at a 
temperature much higher than spin's temperature. Kraus operators, for a qubit, can be written as {\cite{Srikanth_2008}}:
\begin{align}
K_1=\sqrt{p}\begin{pmatrix} 1 & 0 \\ 0 & \sqrt{1-\gamma}
\end{pmatrix}, \ \ K_2=\sqrt{p}\begin{pmatrix} 0 & \sqrt{\gamma} \\ 0 & 0
\end{pmatrix}, \ \ K_3=\sqrt{1-p}\begin{pmatrix} \sqrt{1-\gamma} & 0 \\ 0 & 1
\end{pmatrix},\ \ K_4=\sqrt{1-p}\begin{pmatrix} 0 & 0 \\ \sqrt{\gamma} & 0
\end{pmatrix} .
    \label{eq:Kraus_GAD}
\end{align}

where $p,\lambda\in[0,1]$, $\lambda$ is interpreted as the probability of losing a photon, while $p$ stands for the stationary state probability; $\rho_{\infty}=\text{diag}\{p,1-p\}$. Again, in the case of two qubits, tensor product combination is required to generate the 16 Kraus operators.

\subsubsection{Stochastic dephasing channel (SDC)} describes relaxation of a couple of two-level emitters interacting with a stochastic field \cite{wei2018}. For some quantum states, it has been shown frozen entanglement {\cite{Liu_2016}} and sudden death/birth of discord \cite{wei2018} under the effects of this sort of noise. The Kraus operators associated to this noise are {\cite{Yu_2003}}:
\begin{align}
K_1=\begin{pmatrix}
\sqrt{1-p} & 0 & 0 & 0\\
0 & 1 & 0 & 0\\
0 & 0 & 1 & 0\\
0 & 0 & 0 & \sqrt{1-p}
\end{pmatrix},\ \ K_2=\begin{pmatrix}
\sqrt{p} & 0 & 0 & 0\\
0 & 0 & 0 & 0\\
0 & 0 & 0 & 0\\
0 & 0 & 0 & -\sqrt{p}(1-p)
\end{pmatrix},\ \ K_3=\begin{pmatrix}
0 & 0 & 0 & 0\\
0 & 0 & 0 & 0\\
0 & 0 & 0 & 0\\
0 & 0 & 0 & p\sqrt{2-p}
\end{pmatrix}.
    \label{eq:Kraus_SDC}
\end{align}

Parameters involved in the distinct noisy processes are usually related to the properties of the physical system. For instance, in the case of stochastic  dephasing, parameter $p$ is associated to the phase relaxation time through $p=1-e^{-t/T}$. Thus, evolution from $t=0$ to $t\rightarrow\infty$ is encoded in $p=0$ to $p=1$, respectively.

\subsection{Swapping process}

Entanglement swapping is a way of creating entanglement between a pair of particles that come from different sources and have never interacted with each other in the past. The general protocol was initially introduced in {\cite{Zukowski_1993}} and verified experimentally with photons in {\cite{Zeilinger_1998}}. Similarly, individual photons have been entangled across a distance of 143 km by means of entanglement swapping {\cite{Herbst_2015}}. In terms of quantum correlations, CHSH Bell-nonlocality has been demonstrated to be activated by means of entanglement swapping \cite{W_jcik_2010}. The swapping process works as follows; we consider two pairs of qubits in the states $\rho_{12}$ and $\rho_{34}$, from different sources $S_{12}$ and $S_{34}$, respectively. Qubits 1 and 4 are sent far away from each other and from us, while qubits 2 and 3 keep at hand. Then, we perform a joint Bell measurement on $S_{23}$. Hence, after the measurement, we ask for the amount of quantum resources; steering, entanglement, and discord shared between qubits 1 and 4. The post-measurement state $\rho_{14}$ is computed as: 
\begin{align}
 \rho_{14}=\mathrm{Tr}_{23}\hspace{2pt}\left[\frac{M_{i}\hspace{2pt}\rho_{1234}\hspace{2pt}M_{i}^{\dagger}}{\mathrm{Tr}\hspace{2pt}
[M_{i}\hspace{2pt}\rho_{1234}\hspace{2pt}M_{i}^{\dagger}]}\right].
    \label{eq:state_swap}
\end{align}
where $M_{i}=\mathbb{1}_{2}\otimes \ketbra{\phi_{i}}{\phi_{i}}\otimes \mathbb{1}_{2}$ and $\ket{\phi_{i}}$ stand for the four Bell states: $\ket{\phi_1}=\frac{1}{\sqrt{2}}(\ket{00}+\ket{11})$, $\ket{\phi_2}=\frac{1}{\sqrt{2}}(\ket{00}-\ket{11})$, $\ket{\phi_3}=\frac{1}{\sqrt{2}}(\ket{01}+\ket{10})$ and $\ket{\phi_4}=\frac{1}{\sqrt{2}}(\ket{01}-\ket{10})$.

\section{Results}
\label{results}  
\begin{figure*}[htb]
    \centering
    \includegraphics[width=1\textwidth]{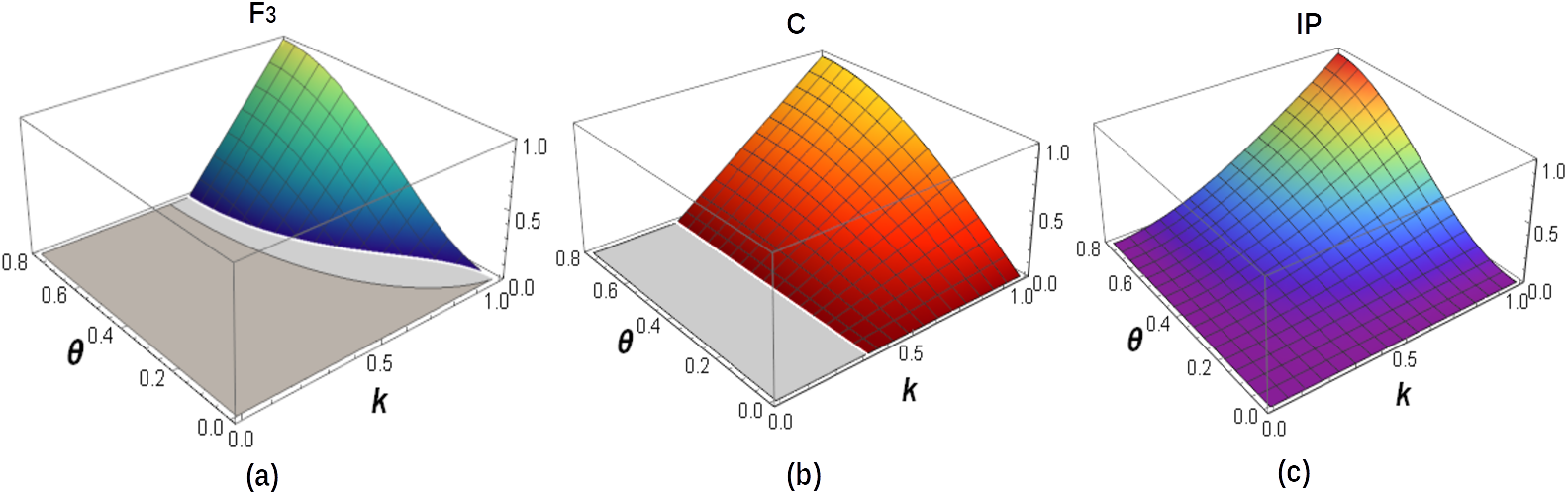}
    \caption{(a) EPR-steering, (b) entanglement and  (c) discord as the interferometric power of quantum states as functions of $\theta$ and $k$ for Almeida et. al. states. Dark gray zone in (a) represents states that obey a LHS model according to Eq. \eqref{eq:uns1}.}
    \label{fig:state_initial}
\end{figure*}

As the starting point of our discussion, we describe the behaviour of EPR-steering, entanglement and discord as function of the two parameters involved in the Almeida et. al. states in Eq. \eqref{eq:state}.

As can be seen in \autoref {fig:state_initial}, almost all Almeida et. al. states have quantum discord, this property only disappears for the extreme lines at $k=0$ and $\theta=0$. However, for entanglement, Almeida et. al. states are divided in two regions; entangled states for $k>1/3$ and arbitrary $\theta$, and separable states with $k\leq1/3$ and any $\theta$. It is worth noting here that, as expected, $\theta$ does not determine the frontier between separable and entangled states, but only the mixture parameter $k$ does. EPR-steering, the more restrictive amongst these quantum properties, is also separated in two zones for the considered states; unsteerable states (gray plateau) and states violating the $F_3$-inequality and therefore displaying EPR-steering. In this case, the frontier is determined by a function between $\theta$ and $k$. It can be compared with the condition for unsteerable states given in Eq. \eqref{eq:uns1}. It is worth bearing in mind that EPR-steering ($F_3$) determines the amount of steering resource a quantum state has. However, condition $F_3=0$ does not imply that the state obeys a LHS model. This is illustrated in \autoref{fig:state_initial}(a), where the dark gray zone on plane $F_3=0$ represents the unsteerability criterion \eqref{eq:uns1}, and it can readily be seen that it does not match the $F_3$-inequality violation frontier given by the greenish surface. In other words, states in the light gray zone do not satisfy the unsteerability criterion, but neither are caught by the violation of the $F_3$-inequality for EPR-steering. Either proving unsteerability or demonstrating the maximal violation of EPR-steering inequalities for quantum states continues to be a very active area of research \cite{LHV2018, betterlocal}. We now analyse the effect of various quantum processes on the amount of EPR-steering, discord, and entanglement that the states possess. We start by analysing the effect of selected noisy channels.

\subsection{Phase damping (PD) noise effects}
\begin{figure*}[htb]
    \centering
    \includegraphics[width=1\textwidth]{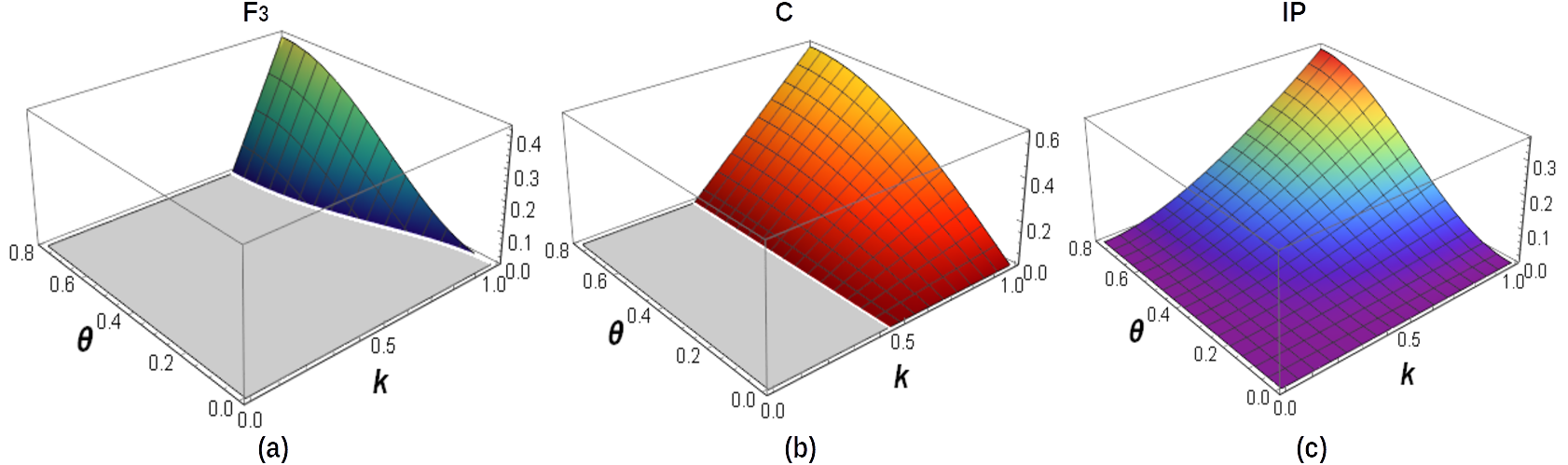}
    \caption{(a) EPR-steering, (b) entanglement and  (c) interferometric power under PD noise. Noisy parameter $p=0.4$. Despite the similar behaviour of the three properties with respect to the corresponding ones in Fig. \ref{fig:state_initial}, all of them achieve lower values due to the noise.}
    \label{fig:PD_state_1}
\end{figure*}

\begin{figure*}[htb]
    \centering
    \includegraphics[width=1\textwidth]{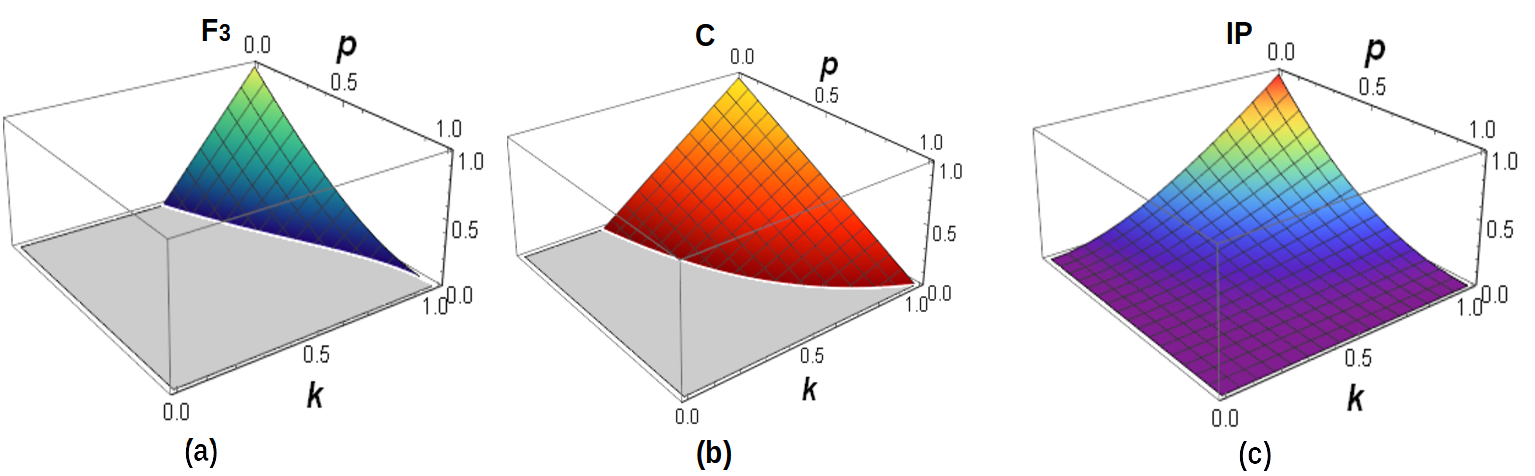}
    \caption{(a) EPR-steering, (b) entanglement and  (c) interferometric power under PD for $\theta=\pi/4$. In terms of the noisy parameter, the fact that steering and entanglement decay at finite values of $p<1$ is interpreted as the phenomenon of sudden death.}
    \label{fig:PD_2}
\end{figure*}

Phase damping noise effects \eqref{eq:Kraus_PD} on Almeida et. al. states  \eqref{eq:state} are  illustrated for a fixed value of the noisy parameter $p$ (\autoref{fig:PD_state_1}) and as a function of $p$ and mixture parameter $k$ (\autoref{fig:PD_2}). 

For a direct comparison with \autoref{fig:state_initial}, the behaviour of quantum properties is shown in \autoref{fig:PD_state_1} in terms of Almeida et. al. state parameters, for the particular noisy case of $p=0.4$. Quantum steering (entanglement) is affected in such a way that its gray zone increases, i.e., some states become unsteerable (separable) due to the effect of the noise. On the other hand, discord continues being presented in all the range of parameters, but decaying a bit faster than in \autoref{fig:state_initial}. Furthermore, it is  worth pointing out that the maximum value achieved by the three properties is considerably lower than the corresponding value in Fig. \ref{fig:state_initial}.

On the other hand, when parameter $\theta$ is fixed, the particular phenomenon of {\it correlation sudden death} is exhibited by EPR-steering (panel \textbf{(a)}) and entanglement (panel \textbf{(b)}), as shown in \autoref{fig:PD_2}. The profile of both quantities is such that when $p$ increases to 1, they decay to zero at a finite value of $p<1$. 
Entanglement is more robust against the sudden death than EPR-steering, with the sudden death taking place for fixed $\theta$ as a function of both $k$ and $p$. However, for fixed $k$, the phenomenon occurs at the same value of $\theta$ (see \autoref{fig:PD_1} in Appendix \ref{app:morecorr}).
Quantum discord decays faster but does not display the sudden death phenomenon.

\subsection{Generalised amplitude damping (GAD) noise effects}
\begin{figure*}[htb]
    \centering
    \includegraphics[width=0.8\textwidth]{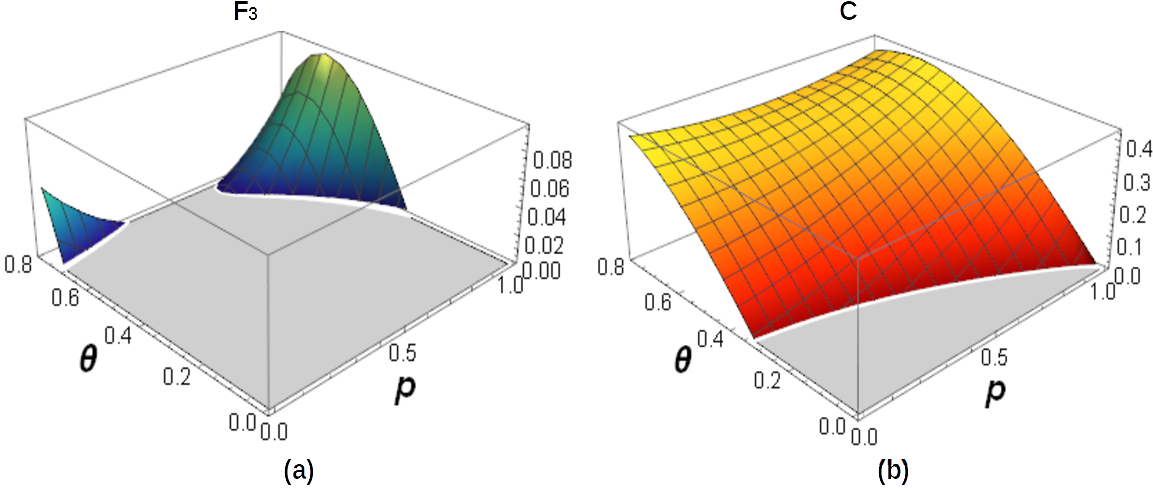}
    \caption{(a) EPR-steering and (b) entanglement under GAD with  $k=0.9$ and $\gamma=0.3$. Another unusual phenomenon named death and revival appears for both properties at different noisy configurations (distinct noisy parameter values).}
    \label{fig:GAD_revival}
\end{figure*}

Generalised amplitude damping (Eq. \eqref{eq:Kraus_GAD}) also leads correlations to decrease and remains quantum discord to be the most robust correlation. However, the behaviour of EPR-steering, as shown in \autoref{fig:GAD_revival} (panel \textbf{(a)}) for $k=0.9$ and $\gamma=0.3$, can be interpreted as a recovering behaviour, or, as usually referred to as originally for entanglement; a death and revival phenomenon. Although the amount of steering in this case is low, it is possible to see a small region (close to $\theta=\pi/4$) where steering decays to zero, but after a $p$-period it increases again. For the set of states here considered, this behaviour occurs for some few configurations, e.g., steering disappears quickly by moving dissipative parameter $\gamma$, while the death and revival phenomenon starts to appear for entanglement (see \autoref{fig:GAD_revival_EN} in Appendix \ref{app:morecorr} for $\gamma=0.6$ and same $k$). Discord is not included in \autoref{fig:GAD_revival} as neither sudden death nor revival phenomena are exhibited by this property.

\subsection{Stochastic dephasing channel (SDC)}
\begin{figure*}[htb]
    \centering
    \includegraphics[width=1\textwidth]{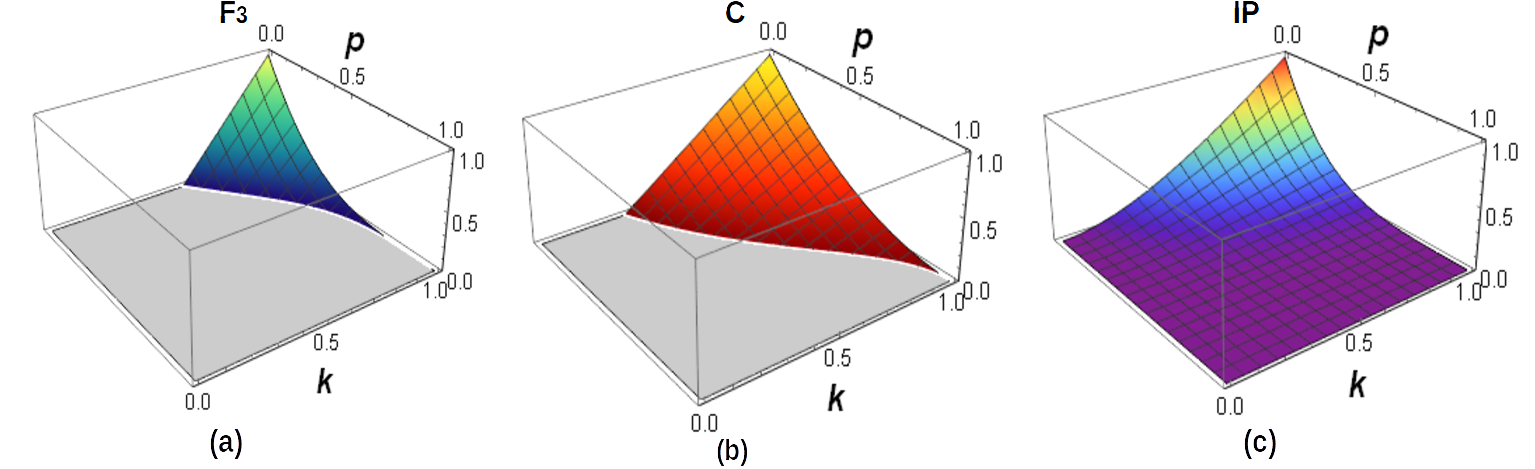}
    \caption{(a) EPR-steering, (b) entanglement and (c) interferometric power under SDC for $\theta=\pi/4$.}
    \label{fig:SDC_1}
\end{figure*}

Stochastic dephasing channel (Eq. \eqref{eq:Kraus_SDC}) evidences decoherent effects similar to the PD noise. \autoref{fig:SDC_1} shows the behaviour of the three properties in terms of initial state parameter $k$ and the SDC parameter $p$. All the properties decay faster than the respective ones shown in \autoref{fig:PD_2} for the PD noise. This can be understood due to the quadratic dependency of the density matrix elements introduced by the SDC noise on parameter $p$ (see Eq. \eqref{eq:SDC_state} and compare with the linear dependency in Eq. \eqref{eq:PD_state}, in Appendix \ref{app:noisyqstates}). Despite the sudden death of discord reported in \cite{wei2018} for some states under SDC noise, we do not find any scenario in which that kind of behaviour takes place for the considered quantum states.

On the other hand, sudden death of steering and entanglement also is revealed by noisy effects due to SDC as shown in panels \textbf{(a)} and \textbf{(b)} of \autoref{fig:SDC_1}.  Similarly, for fixed $k$ the phenomenon occurs for both quantum properties but at different facets; for entanglement, it happens at the same value of $p$ and arbitrary $\theta$, while the phenomenon's profile appears as a function of both parameters for steering (see \autoref{fig:SDC_2} in Appendix \ref{app:morecorr}).

\subsection{Swapping protocol effects}
\begin{figure*}[htb]
    \centering
    \includegraphics[width=1\textwidth]{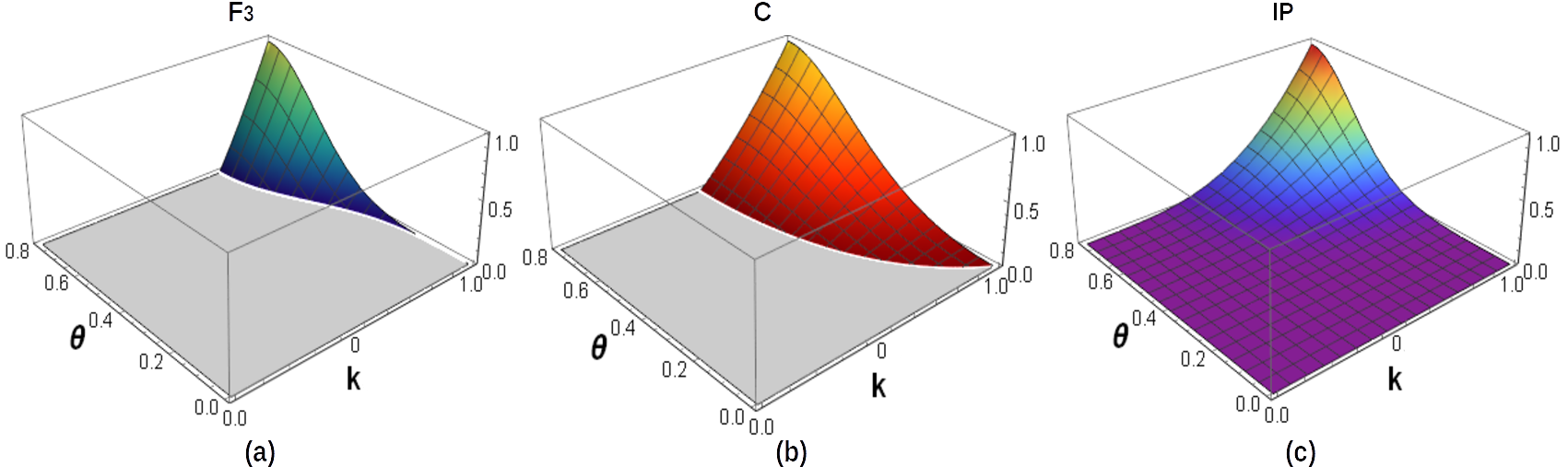}
    \caption{(a) EPR-steering, (b) entanglement and (c) interferometric power of states $\rho_{14}$ after swapping.  Initial states are considered to be the same $\rho_{12}=\rho_{34}=\rho(k,\theta)$ and the projectors defined with $\ket{\phi_{1}}$ and $\ket{\phi_{2}}$.}
    \label{fig:state_swap_1}
\end{figure*}

\begin{figure*}[htb]
    \centering
    \includegraphics[width=1\textwidth]{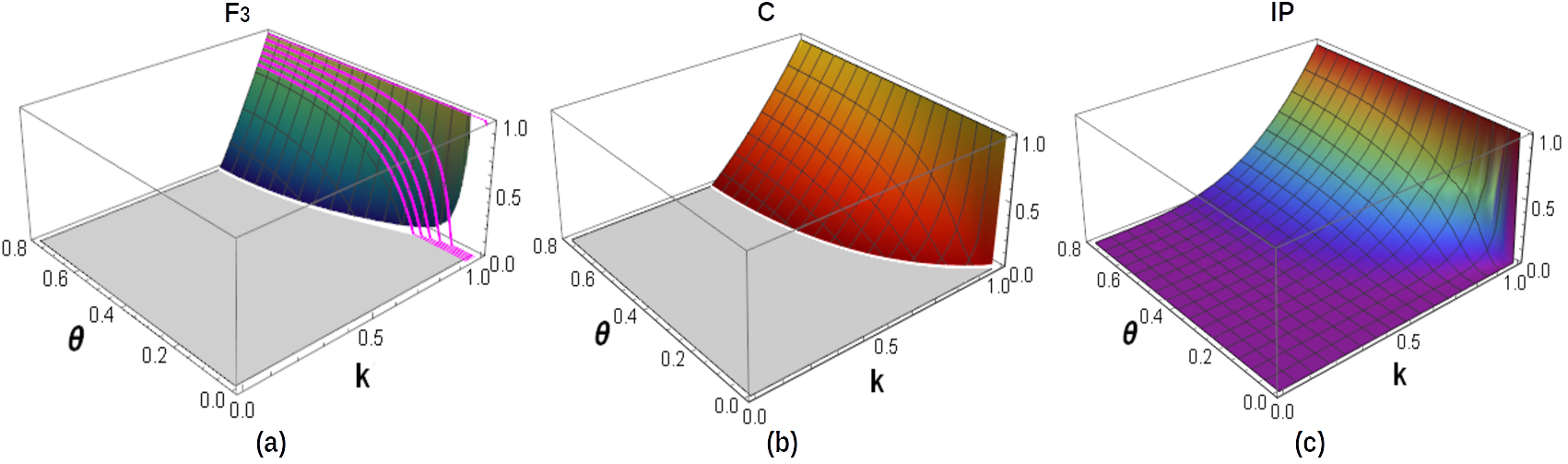}
    \caption{(a) EPR-steering, (b) entanglement and (c) interferometric power  for states $\rho_{14}$ after swapping. Initial states $\rho_{12}=\rho_{34}=\rho(k,\theta)$ and  projectors defined by means of $\ket{\phi_{3}}$ and $\ket{\phi_{4}}$.}
    \label{fig:state_swap_2}
\end{figure*}

Swapping protocol effects on the three quantum properties here considered can be explored in several distinct setups by tailoring the considered states for sources $S_{12}$ and $S_{34}$, and the set of measurements applied on the pair in $S_{23}$. We consider the following particular setup: $\rho_{12}=\rho_{34}= \rho(k,\theta)$ and measurements on $S_{23}$ to be Bell measurements as in Eq. \eqref{eq:state_swap}. EPR-steering, entanglement and discord are shown in terms of the initial state parameters. \autoref{fig:state_swap_1} considers the projectors $M_{1,2}$, and \autoref{fig:state_swap_2} the projectors $M_{3,4}$.

First, we analyse the case for \autoref{fig:state_swap_1} where projectors $M_{1,2}$ are implemented. We start by highlighting that the swapping protocol allows for completely non-correlated pairs of particles to become correlated. In \autoref{fig:state_swap_1}, even though the amount of correlations being generated are less than the amount shared by the initial states $\rho_{12}=\rho_{34}=\rho(k,\theta)$, they are still strictly different than zero for some region. This makes the effects of swapping protocol to stand in contrast difference to the case from the noisy processes. In short, that despite the behaviour of the three quantum properties of the system in $S_{14}$, the system in general benefits from the swapping process as a whole.

A natural follow up question to ask is then, how much correlated could the final system $\rho_{14}$ become? In \autoref{fig:state_swap_2} we report a scenario where the correlations of the final state $\rho_{14}$ surpass the ones of the initial states $\rho_{12}=\rho_{34}=\rho(k,\theta)$. 
In \autoref{fig:state_swap_2}, the case for $k=1$ and arbitrary $\theta$ displays a scenario where the final state is the maximally entangled state of two qubits (the singlet state up to local unitaries). This is a bit more general result than the original entanglement swapping protocol that transfers maximally entangled states to maximally entangled state, and it is in complete agreement with some results reported in \cite{Zukowski_1993, Kirby_2016}.

More interestingly, in \autoref{fig:state_swap_2}, the general case of $k\neq1$ evidences that the amount of swapped quantum properties can be larger than the shared by the initial states. In particular, when $k\rightarrow1$, all the quantum properties  quickly increase to a value very close to $1$ as functions of $\theta$ (to better illustrate this behaviour, we have added some 2D curves on the steering's surface for values from $k=0.95$ to $1$ - panel \textbf{(a)}, with a similar behaviour being exhibited by entanglement and discord). 
Hence, comparing \autoref{fig:state_swap_1} and \autoref{fig:state_swap_2}, we have that Bell's projectors $M_3$ and $M_4$ allow for the generation of larger amount of quantum properties than the Bell's projectors $M_1$ and $M_2$, and than the amount of the respective properties in the initial states $\rho_{12}$ and $\rho_{34}$ (compare with \autoref{fig:state_initial}). This therefore highlights the importance of the Bell measurement to be implemented and particularly, the post-selected state to be chosen after the swapping process. In this case, projectors $M_3$ and $M_4$ are useful (as there is an increment on the amount of correlations), whilst $M_1$ and $M_2$ are not useful (since they generally degrade correlations).
\begin{figure*}[htb]
    \centering
    \includegraphics[width=0.8\textwidth]{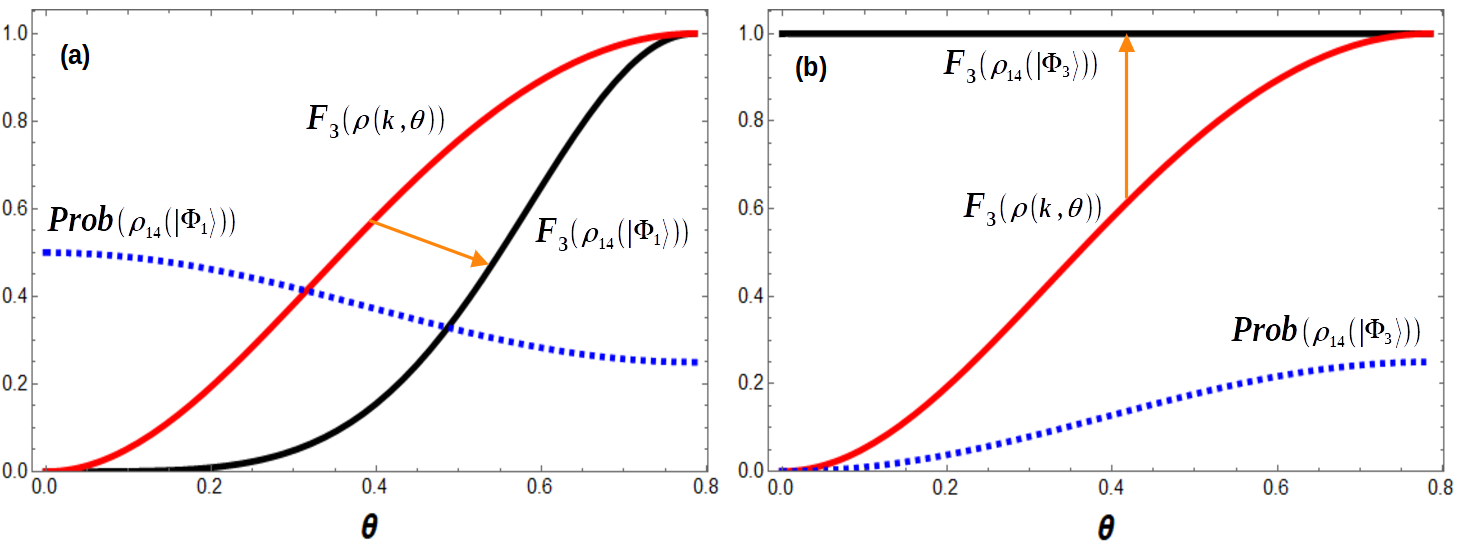}
    \caption{EPR-steering of initial state $\rho(k,\theta)$ (red), final state $\rho_{14}$ (black), and swapping probability $Prob(\rho_{14})$ (dotted blue), as functions of $\theta$. (a) Swapping using the projector
    $\ketbra{\phi_{1}}{\phi_{1}}$, and (b) the projector $\ketbra{\phi_{3}}{\phi_{3}}$. with $k=1$.}
    \label{fig:steering_2d_2}
\end{figure*}

As in the general case of partially-correlated mixed states, the post-measurement quantum properties in the swapping protocol exhibit distinct behaviour depending on the implemented Bell's projector, it is important to address the probability of occurrence of each result. In particular, for EPR-steering, we compare the effects of projecting on $\ket{\phi_1}$ and $\ket{\phi_3}$. In \autoref{fig:steering_2d_2}, we have EPR-steering of the initial state (red), final state (black), and swapping probability (dotted blue), for $k=1$ and arbitrary $\theta$. In \autoref{fig:steering_2d_2} \textbf{(a)} we address the projector $\phi_1$, and we have that the swapping process generally degrades the correlations. In \autoref{fig:steering_2d_2} \textbf{(b)} we address the projector $\phi_3$ and the swapping process happen to further increase the amount of EPR-steering, taking it to the maximum possible amount for all $\theta$. This however, at the expense of the swapping probability decreasing and going to zero when $\theta\rightarrow0$. These results evidence the existence of a trade-off between the extractable amount of quantum properties (EPR-steering in this case) and the swapping probability.
\begin{figure*}[htb]
    \centering
    \includegraphics[width=0.8\textwidth]{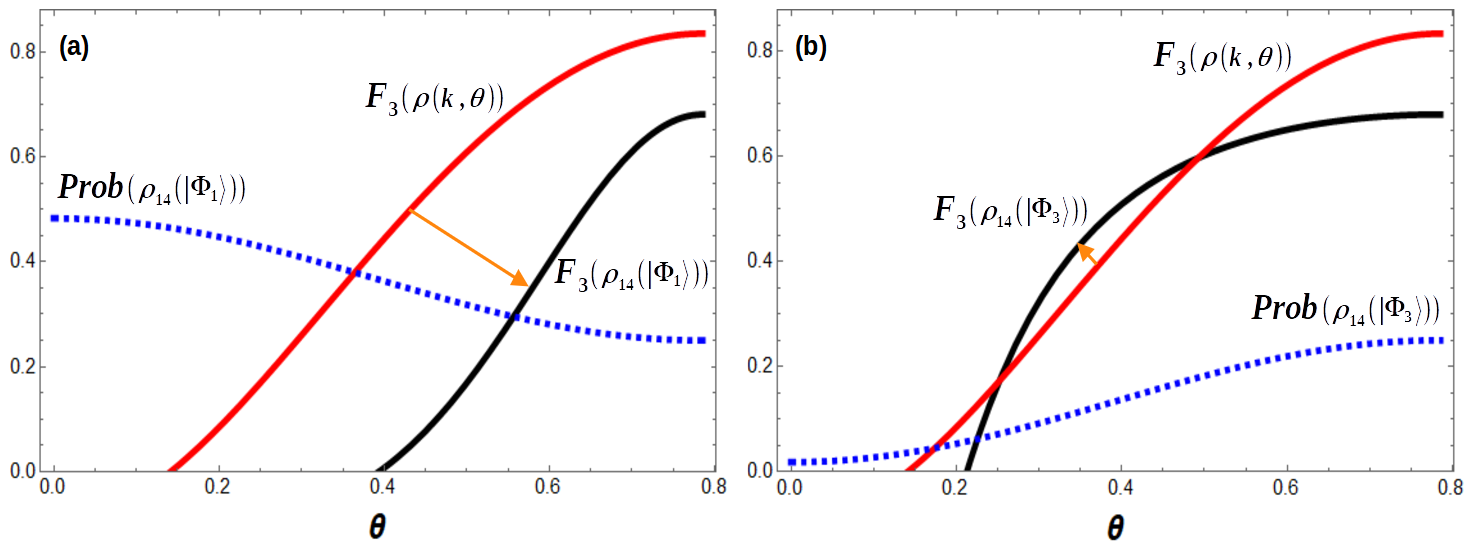}
    \caption{EPR-steering of initial state $\rho(k,\theta)$ (red), final state $\rho_{14}$ (black), and swapping probability $Prob(\rho_{14})$ (dotted blue), as functions of $\theta$. (a) Swapping using the projector
    $\ketbra{\phi_{1}}{\phi_{1}}$, and (b) the projector $\ketbra{\phi_{3}}{\phi_{3}}$. with $k=0.93$.}
    \label{fig:steering_2d_1}
\end{figure*}

In \autoref{fig:steering_2d_1}, we have the same scenario but with $k=0.93$, i.e., initial mixed states, and arbitrary $\theta$. As shown in \autoref{fig:steering_2d_1} \textbf{(a)}, projector $\phi_1$ degrades the quantum property of EPR-steering, and in contrast to the result in \autoref{fig:steering_2d_2} \textbf{(a)}, some initial states are swapped to unsteerable states ($\theta<0.4$). In \autoref{fig:steering_2d_1} \textbf{(b)} we address the projector $\phi_3$ and although pretty narrow, there  is still a zone where the amount of EPR-steering enhances with respect to the shared one by the initial states. Despite the trade-off between the amount of correlations and the probability of occurrence, it is worth noting that the probability of having after-swapping correlations larger than initial correlations, continues to be significant ($\sim10-15\%$), thus this correlation enhancement being  feasible to be detected in experimental setups. 

A more elaborated discussion about swapping scenarios as well as local filtering effects on quantum non-nonlocality, steering, usefulness for teleportation and the quantum obesity \cite{Milne_2014, Jevtic_2014} is part of an upcoming work \cite{rosario2021}.

\section{Conclusions}
\label{conclusion}
We have discussed in detail the behaviour of three quantum properties which are of relevance for nowadays quantum technologies; EPR-steering, entanglement and quantum discord. For doing so, we have considered two distinct processes on a set of two-qubit states; on the first hand, we have explored these quantum properties under three different types of noisy channels. Particularly, we have taken into account effects of phase damping, generalised amplitude damping and stochastic dephasing channel. On the second hand, we have also explored the protocol of entanglement swapping to analyse the amount of quantum properties that can be transferred to two spatially-separated qubits.

Regarding the noisy processes, we have evidenced that both dephasing noises produce similar losses on the three quantum properties. In particular, EPR-steering and entanglement exhibit the phenomenon of \emph{sudden death} which appears at an earlier stage for EPR-steering. Comparing both noises, SDC is more detrimental than PD. On the other hand, a richer behaviour is shown for the generalised amplitude damping (GAD), where the phenomenon referred to as \emph{death and revival} is also displayed by EPR-steering and entanglement. We highlight that this phenomenon does not take place for the two previous noises. These results therefore evidence that, whilst the noisy channels here considered generally reduce the amount of properties of the system, this happens in a non-trivial manner for some scenarios, as evidenced by the \emph{sudden death} as well as the \emph{death and revival} of EPR-steering and entanglement induced by the GAD.

Additionally, the process of transferring quantum properties by means of swapping is also discussed for the three aforementioned quantum properties. Although the behaviour of all of them is such that the initial amount of quantum properties for $\rho_{12}=\rho_{34}$ is not completely transmitted to the final state $\rho_{14}$, it is worth noting that this process can always be seen as a generator of quantum properties, as the parties $1$ and $4$ are initially independent to each other and therefore no correlated whatsoever. Another important point is that after swapping process has taken place, the final state of $\rho_{14}$ can display a different behaviour depending on the Bell's projector being used in the intermediate measurement. Interestingly, and in sharp contrast with the previous scenarios with noisy channels, the process of swapping can \emph{enhance} the three quantum properties here considered in a range close to $k\rightarrow1$, meaning that it allows for more quantum information to be shared, between parties $\rho_{14}$, than the initially shared by parties $\rho_{12}$ and $\rho_{34}$. Furthermore, we analyse a scenario where the swapping process allows for the generation of the maximal possible amount of correlations, meaning that initial partially correlated states can be swapped to final states that maximally display EPR-steering as well as entanglement, meaning the final state is the single state, up to local unitary transformations. Interestingly however, this maximal extraction is effectively limited by the swapping probability with which the phenomenon can take place and consequently, evidencing a trade-off between the transferring of these quantum properties and the probability with which this can happen. A more elaborated discussion with more general entanglement swapping and various other quantum properties is a part of an upcoming work \cite{rosario2021}.

\section*{Acknowledgement}
A.F.D. acknowledges financial support from COLCIENCIAS (Grant 756-2016) and the UK EPSRC (EP/L015730/1). P.R.V and C.E.S. acknowledge funding from Universidad de C\'ordoba (Grant FCB-08-19). C.E.S also acknowledges the Colombian Science, Technology and Innovation Fund--General Royalties System (Fondo CTeI---Sistema General de Regal\'ias) and Gobernaci\'on del Valle del Cauca (Grant BPIN 2013000100007).

\bibliographystyle{unsrt}

\bibliography{cas-refs.bib}

\begin{thebibliography}{10}

\bibitem{pramanik2019}
Tanumoy Pramanik, Young-Wook Cho, Sang-Wook Han, Sang-Yun Lee, Sung Moon, and
  Yong-Su Kim.
\newblock Nonlocal quantum correlations under amplitude damping decoherence.
\newblock {\em Phys. Rev. A}, 100:042311, Oct 2019.

\bibitem{wang2021}
Guoming Wang, Dax~Enshan Koh, Peter~D. Johnson, and Yudong Cao.
\newblock Minimizing estimation runtime on noisy quantum computers.
\newblock {\em PRX Quantum}, 2:010346, Mar 2021.

\bibitem{chitambar2019}
Eric Chitambar and Gilad Gour.
\newblock Quantum resource theories.
\newblock {\em Rev. Mod. Phys.}, 91:025001, Apr 2019.

\bibitem{garding2021}
Elias Riedel~Gårding, Nicolas Schwaller, Chun~Lam Chan, Su~Yeon Chang, Samuel
  Bosch, Frederic Gessler, Willy~Robert Laborde, Javier~Naya Hernandez, Xinyu
  Si, Marc-André Dupertuis, and Nicolas Macris.
\newblock Bell diagonal and werner state generation: Entanglement,
  non-locality, steering and discord on the ibm quantum computer.
\newblock {\em Entropy}, 23(7), 2021.

\bibitem{Su2021}
Daiqin Su, Robert Israel, Kunal Sharma, Haoyu Qi, Ish Dhand, and Kamil
  Br{\'{a}}dler.
\newblock Error mitigation on a near-term quantum photonic device.
\newblock {\em {Quantum}}, 5:452, May 2021.

\bibitem{huang2021}
Chang-Jiang Huang, Guo-Yong Xiang, Yu~Guo, Kang-Da Wu, Bi-Heng Liu, Chuan-Feng
  Li, Guang-Can Guo, and Armin Tavakoli.
\newblock Nonlocality, steering, and quantum state tomography in a single
  experiment.
\newblock {\em Phys. Rev. Lett.}, 127:020401, Jul 2021.

\bibitem{zhao2020}
Yuan-Yuan Zhao, Huan-Yu Ku, Shin-Liang Chen, Hong-Bin Chen, Franco Nori,
  Guo-Yong Xiang, Chuan-Feng Li, Guang-Can Guo, and Yueh-Nan Chen.
\newblock Experimental demonstration of measurement-device-independent measure
  of quantum steering.
\newblock {\em npj Quantum Information}, 6:77:2056--6387, Sep 2020.

\bibitem{bell1964}
J.~S. Bell.
\newblock On the einstein podolsky rosen paradox.
\newblock {\em Physics Physique Fizika}, 1:195--200, Nov 1964.

\bibitem{wiseman2007}
H.~M. Wiseman, S.~J. Jones, and A.~C. Doherty.
\newblock Steering, entanglement, nonlocality, and the einstein-podolsky-rosen
  paradox.
\newblock {\em Phys. Rev. Lett.}, 98:140402, Apr 2007.

\bibitem{branciard2012}
Cyril Branciard, Eric~G. Cavalcanti, Stephen~P. Walborn, Valerio Scarani, and
  Howard~M. Wiseman.
\newblock One-sided device-independent quantum key distribution: Security,
  feasibility, and the connection with steering.
\newblock {\em Phys. Rev. A}, 85:010301, Jan 2012.

\bibitem{Horodecki_1996}
Ryszard Horodecki, Michał Horodecki, and Paweł Horodecki.
\newblock Teleportation, bell’s inequalities and inseparability.
\newblock {\em Physics Letters A}, 222(1-2):21–25, Oct 1996.

\bibitem{zurek2001}
Harold Ollivier and Wojciech~H. Zurek.
\newblock Quantum discord: A measure of the quantumness of correlations.
\newblock {\em Phys. Rev. Lett.}, 88:017901, Dec 2001.

\bibitem{ferraro2010}
A.~Ferraro, L.~Aolita, D.~Cavalcanti, F.~M. Cucchietti, and A.~Ac\'{\i}n.
\newblock Almost all quantum states have nonclassical correlations.
\newblock {\em Phys. Rev. A}, 81:052318, May 2010.

\bibitem{adesso2016}
Gerardo Adesso, Thomas~R Bromley, and Marco Cianciaruso.
\newblock Measures and applications of quantum correlations.
\newblock {\em Journal of Physics A: Mathematical and Theoretical},
  49(47):473001, Nov 2016.

\bibitem{Milne_2014}
Antony Milne, Sania Jevtic, David Jennings, Howard Wiseman, and Terry Rudolph.
\newblock Quantum steering ellipsoids, extremal physical states and monogamy.
\newblock {\em New Journal of Physics}, 16(8):083017, Aug 2014.

\bibitem{Jevtic_2014}
Sania Jevtic, Matthew Pusey, David Jennings, and Terry Rudolph.
\newblock Quantum steering ellipsoids.
\newblock {\em Physical Review Letters}, 113(2), Jul 2014.

\bibitem{ducuara2020sudden}
Andrés~F. Ducuara, Cristian~E. Susa, and John~H. Reina.
\newblock Emergence of maximal hidden quantum correlations and its trade-off
  with the filtering probability in dissipative two-qubit systems, 2021.

\bibitem{yadin2021}
Benjamin Yadin, Matteo Fadel, and Manuel Gessner.
\newblock Metrological complementarity reveals the einstein-podolsky-rosen
  paradox.
\newblock {\em Nature Communications}, 12(2410), 2021.

\bibitem{yang2021}
Huan Yang, Zhi-Yong Ding, Xue-Ke Song, Hao Yuan, Dong Wang, Jie Yang, Chang-Jin
  Zhang, and Liu Ye.
\newblock Verification of complementarity relations between quantum steering
  criteria using an optical system.
\newblock {\em Phys. Rev. A}, 103:022207, Feb 2021.

\bibitem{bart2016}
Karol Bartkiewicz, Antonín Cernoch, Karel Lemr, Adam Miranowicz, and Franco
  Nori.
\newblock Experimental temporal quantum steering.
\newblock {\em Scientific Reports}, 6:38076, Feb 2016.

\bibitem{xu2017}
JS. Xu, Li~CF., and Guo GC.
\newblock Experimental investigation of the dynamics of quantum discord in
  optical systems. in: Fanchini f., soares pinto d., adesso g. (eds) lectures
  on general quantum correlations and their applications.
\newblock {\em Quantum Science and Technology}, 6:473--484, Feb 2017.

\bibitem{allati2014}
A.~El, Allati, S.~Robles-Pérez, and M.~El, Baz.
\newblock Quantum discord in optical coherent states.
\newblock {\em International Journal of Theoretical Physics}, 53:38076, Feb
  2014.

\bibitem{costa2016}
A.~C.~S. Costa and R.~M. Angelo.
\newblock Quantification of einstein-podolski-rosen steering for two-qubit
  states.
\newblock {\em Physical Review A}, 93(2), Feb 2016.

\bibitem{girolami2014}
Davide Girolami, Alexandre~M. Souza, Vittorio Giovannetti, Tommaso Tufarelli,
  Jefferson~G. Filgueiras, Roberto~S. Sarthour, Diogo~O. Soares-Pinto, Ivan~S.
  Oliveira, and Gerardo Adesso.
\newblock Quantum discord determines the interferometric power of quantum
  states.
\newblock {\em Phys. Rev. Lett.}, 112:210401, May 2014.

\bibitem{Wootters_1998}
William~K. Wootters.
\newblock Entanglement of formation of an arbitrary state of two qubits.
\newblock {\em Physical Review Letters}, 80(10):2245–2248, Mar 1998.

\bibitem{cavalcanti2009}
E.~G. Cavalcanti, S.~J. Jones, H.~M. Wiseman, and M.~D. Reid.
\newblock Experimental criteria for steering and the einstein-podolsky-rosen
  paradox.
\newblock {\em Physical Review A}, 80(3), Sep 2009.

\bibitem{vedral2012}
Kavan Modi, Aharon Brodutch, Hugo Cable, Tomasz Paterek, and Vlatko Vedral.
\newblock The classical-quantum boundary for correlations: Discord and related
  measures.
\newblock {\em Rev. Mod. Phys.}, 84:1655--1707, Nov 2012.

\bibitem{bera2017}
Anindita Bera, Tamoghna Das, Debasis Sadhukhan, Sudipto~Singha Roy, Aditi
  Sen(De), and Ujjwal Sen.
\newblock Quantum discord and its allies: a review of recent progress.
\newblock 81(2):024001, dec 2017.

\bibitem{girolami2011}
Davide Girolami and Gerardo Adesso.
\newblock Quantum discord for general two-qubit states: Analytical progress.
\newblock {\em Physical Review A}, 83(5), May 2011.

\bibitem{davidovich2010}
A.~Auyuanet and L.~Davidovich.
\newblock Quantum correlations as precursors of entanglement.
\newblock {\em Physical Review A}, 82(3), Sep 2010.

\bibitem{Braunstein_1994}
Samuel~L. Braunstein and Carlton~M. Caves.
\newblock Statistical distance and the geometry of quantum states.
\newblock {\em Phys. Rev. Lett.}, 72:3439--3443, May 1994.

\bibitem{Braun_2018}
Daniel Braun, Gerardo Adesso, Fabio Benatti, Roberto Floreanini, Ugo Marzolino,
  Morgan~W. Mitchell, and Stefano Pirandola.
\newblock Quantum-enhanced measurements without entanglement.
\newblock {\em Reviews of Modern Physics}, 90(3), Sep 2018.

\bibitem{mafalda2007}
Mafalda~L. Almeida, Stefano Pironio, Jonathan Barrett, Géza Tóth, and Antonio
  Acín.
\newblock Noise robustness of the nonlocality of entangled quantum states.
\newblock {\em Physical Review Letters}, 99(4), Jul 2007.

\bibitem{LHV2018}
Mathieu Fillettaz, Flavien Hirsch, S\'ebastien Designolle, and Nicolas Brunner.
\newblock Algorithmic construction of local models for entangled quantum
  states: Optimization for two-qubit states.
\newblock {\em Phys. Rev. A}, 98:022115, Aug 2018.

\bibitem{Bowles_2016}
Joseph Bowles, Flavien Hirsch, Marco~Túlio Quintino, and Nicolas Brunner.
\newblock Sufficient criterion for guaranteeing that a two-qubit state is
  unsteerable.
\newblock {\em Physical Review A}, 93(2), Feb 2016.

\bibitem{kraus1983}
K.~Kraus.
\newblock {\em States, effects and operations}.
\newblock Springer-Verlag Berlin Heidelberg, 1983.

\bibitem{nielsen2000}
Michael~A. Nielsen and Isaac~L. Chuang.
\newblock {\em Quantum Computation and Quantum Information}.
\newblock Cambridge University Press, 2000.

\bibitem{Srikanth_2008}
R.~Srikanth and Subhashish Banerjee.
\newblock Squeezed generalized amplitude damping channel.
\newblock {\em Physical Review A}, 77(1), Jan 2008.

\bibitem{wei2018}
Wei Xia, Jin-Xing Hou, Wang Xiao-Hui, and Si-Yuan Liu.
\newblock The sudden death and sudden birth of quantum discord.
\newblock {\em Scientific Reports}, 8, 03 2018.

\bibitem{Liu_2016}
Bi-Heng Liu, Xiao-Min Hu, Jiang-Shan Chen, Chao Zhang, Yun-Feng Huang,
  Chuan-Feng Li, Guang-Can Guo, G\"oktu\ifmmode \breve{g}\else~\u{g}\fi{}
  Karpat, Felipe~F. Fanchini, Jyrki Piilo, and Sabrina Maniscalco.
\newblock Time-invariant entanglement and sudden death of nonlocality.
\newblock {\em Phys. Rev. A}, 94:062107, Dec 2016.

\bibitem{Yu_2003}
T.~Yu and J.~H. Eberly.
\newblock Qubit disentanglement and decoherence via dephasing.
\newblock {\em Physical Review B}, 68(16), Oct 2003.

\bibitem{Zukowski_1993}
M.~\ifmmode~\dot{Z}\else \.{Z}\fi{}ukowski, A.~Zeilinger, M.~A. Horne, and
  A.~K. Ekert.
\newblock ``event-ready-detectors'' bell experiment via entanglement swapping.
\newblock {\em Phys. Rev. Lett.}, 71:4287--4290, Dec 1993.

\bibitem{Zeilinger_1998}
Jian-Wei Pan, Dik Bouwmeester, Harald Weinfurter, and Anton Zeilinger.
\newblock Experimental entanglement swapping: Entangling photons that never
  interacted.
\newblock {\em Phys. Rev. Lett.}, 80:3891--3894, May 1998.

\bibitem{Herbst_2015}
Thomas Herbst, Thomas Scheidl, Matthias Fink, Johannes Handsteiner, Bernhard
  Wittmann, Rupert Ursin, and Anton Zeilinger.
\newblock Teleportation of entanglement over 143 km.
\newblock {\em Proceedings of the National Academy of Sciences},
  112(46):14202–14205, Nov 2015.

\bibitem{W_jcik_2010}
Antoni Wójcik, Joanna Modławska, Andrzej Grudka, and Mikołaj Czechlewski.
\newblock Violation of clauser–horne–shimony–holt inequality for states
  resulting from entanglement swapping.
\newblock {\em Physics Letters A}, 374(48):4831–4833, Nov 2010.

\bibitem{betterlocal}
Flavien Hirsch, Marco~T{\'{u}}lio Quintino, Tam{\'{a}}s V{\'{e}}rtesi, Miguel
  Navascu{\'{e}}s, and Nicolas Brunner.
\newblock Better local hidden variable models for two-qubit {W}erner states and
  an upper bound on the {G}rothendieck constant {$K_G(3)$}.
\newblock {\em {Quantum}}, 1:3, April 2017.

\bibitem{Kirby_2016}
Brian~T. Kirby, Siddhartha Santra, Vladimir~S. Malinovsky, and Michael Brodsky.
\newblock Entanglement swapping of two arbitrarily degraded entangled states.
\newblock {\em Physical Review A}, 94(1), Jul 2016.

\bibitem{rosario2021}
Pedro Rosario, Andrés Ducuara, and Cristian Susa.
\newblock Characterising quantum entanglement-based properties and beyond
  through general entanglement swapping and optimal local filtering.
\newblock {\em To be submitted}, 2021.

\end{thebibliography}

\appendix
\section{Quantum states after noisy effects}
\label{app:noisyqstates}
In this Appendix we show the explicit form of states after the action of the three considered noisy processes.

After the action of PD noise, states \eqref{eq:state} are transformed ($ \rho(k,\theta) \rightarrow \rho^{PD}(k,\theta,p)$) as shown in Eq. \eqref{eq:PD_state} (the coherences - non-digonal elements are linearly affected by the noisy parameter $p$).
\begin{eqnarray}
\label{eq:PD_state}
\rho^{PD}(k,\theta,p)=
\begin{pmatrix} 
\frac{(1+k)\cos^{2}{\theta}}{2} &0&0&k(1-p)\cos{\theta}\sin{\theta} \\ 0 & \frac{(1-k)\cos^{2}{\theta}}{2}&0&0\\0&0&\frac{(1-k)\sin^{2}{\theta}}{2}&0 \\k(1-p)\cos{\theta}\sin{\theta}&0&0&\frac{(1+k)\sin^{2}{\theta}}{2}
\end{pmatrix} .
\end{eqnarray}

On the other hand, the transformation given by the GAD noise ($\rho(k,\theta) \rightarrow \rho^{GAD}(k,\theta,p,\gamma)$ is more complex than the dephasing one. In this case, all the density matrix's elements are affected by the noise as shown in Eq. \eqref{eq:GAD_state}.
\begin{eqnarray}
\label{eq:GAD_state}
\rho^{GAD}(k,\theta,p,\gamma)=
\begin{pmatrix} 
M_{1} &0&0&k(1-\gamma)\cos{\theta}\sin{\theta} \\ 0 & M_{2} &0&0\\0&0& M_{3}&0 \\k(1-\gamma)\cos{\theta}\sin{\theta}&0&0& M_{4}
\end{pmatrix} ,
\end{eqnarray}
where $M_{1}=\frac{1}{4}(k(1-\gamma)^{2}+(-1 + \gamma -2p\gamma)^{2}-(1+k)(\gamma-1)(1+(2p-1)\gamma)\cos{(2\theta)})$, $M_{2}=\frac{1}{4}(1-k(1-\gamma)^{2}-(1-2p)^{2}\gamma^{2}+(\gamma-1)(-1+k+(1+k)(-1+2p)\gamma)\cos{(2\theta)})$, $M_{3}=\frac{1}{4}(1-k(1-\gamma)^{2}-(1-2p)^{2}\gamma^{2}+(\gamma-1)(1-k+(1+k)(2p-1)\gamma)\cos{(2\theta)})$ and $M_{4}=\frac{1}{4}(k(1-\gamma)^{2}+(1 + \gamma -2p\gamma)^{2}-(1+k)(\gamma-1)(-1+(2p-1)\gamma)\cos{(2\theta)})$\\ \\
Finally, the transformation throughout the SDC noise ($ \rho(k,\theta) \rightarrow \rho^{SDC}(k,\theta,p)$) is shown in Eq. \eqref{eq:SDC_state}. The effects are quite similar to those of the PD with the difference that the noisy parameter dependence is quadratic.
\begin{eqnarray}
\label{eq:SDC_state}
\rho^{SDC}(k,\theta,p)=
\begin{pmatrix} 
\frac{(1+k)\cos^{2}{\theta}}{2} &0&0&k(1-p)^{2}\cos{\theta}\sin{\theta} \\ 0 & \frac{(1-k)\cos^{2}{\theta}}{2}&0&0\\0&0&\frac{(1-k)\sin^{2}{\theta}}{2}&0 \\k(1-p)^{2}\cos{\theta}\sin{\theta}&0&0&\frac{(1+k)\sin^{2}{\theta}}{2}
\end{pmatrix} .
\end{eqnarray}

\section{Correlations for initial state under noisy PD, GAD and SDC}
\label{app:morecorr}

\begin{figure*}[htb]
\centering
\includegraphics[width=1\textwidth]{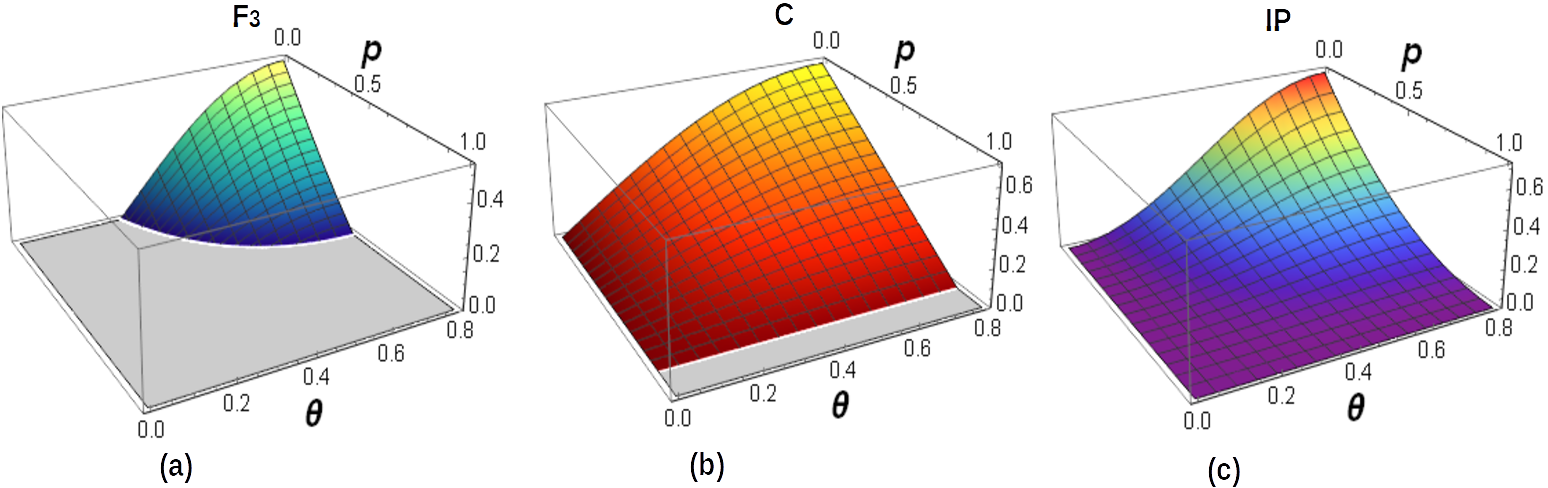}
\caption{(a) EPR-steering, (b) entanglement and  (c) interferometric power under PD noise. fixed $k=4/5$.}
\label{fig:PD_1}
\end{figure*}

\begin{figure*}[htb]
\centering
\includegraphics[width=0.8\textwidth]{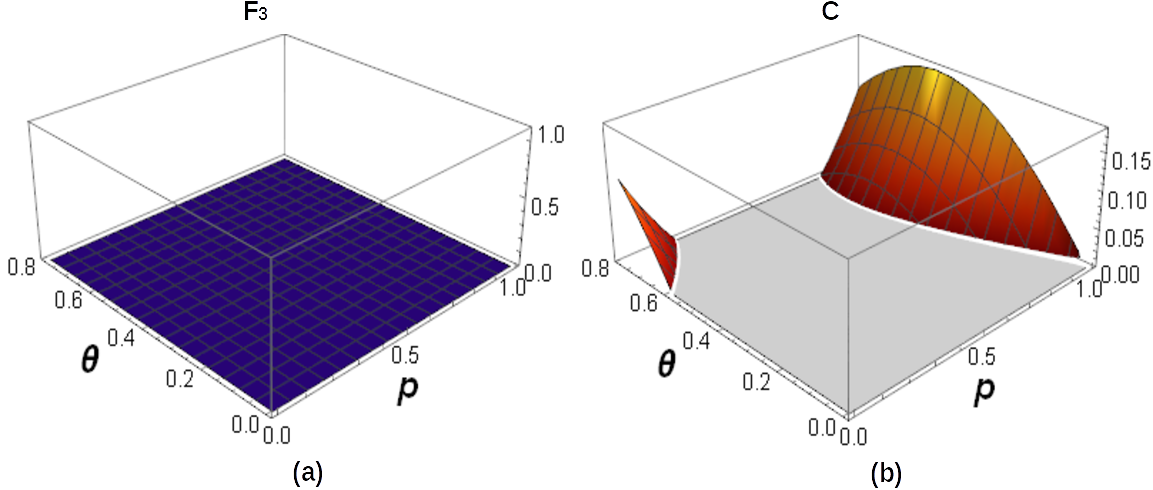}
\caption{(a) EPR-steering and (b) entanglement under GAD with  $k=0.9$ and $\gamma=0.6$. Comparing to Fig. \ref{fig:GAD_revival} in the main text, the revival phenomenon in this case is exhibited by entanglement instead.}
\label{fig:GAD_revival_EN}
\end{figure*}

\begin{figure*}[htb]
\centering
\includegraphics[width=1\textwidth]{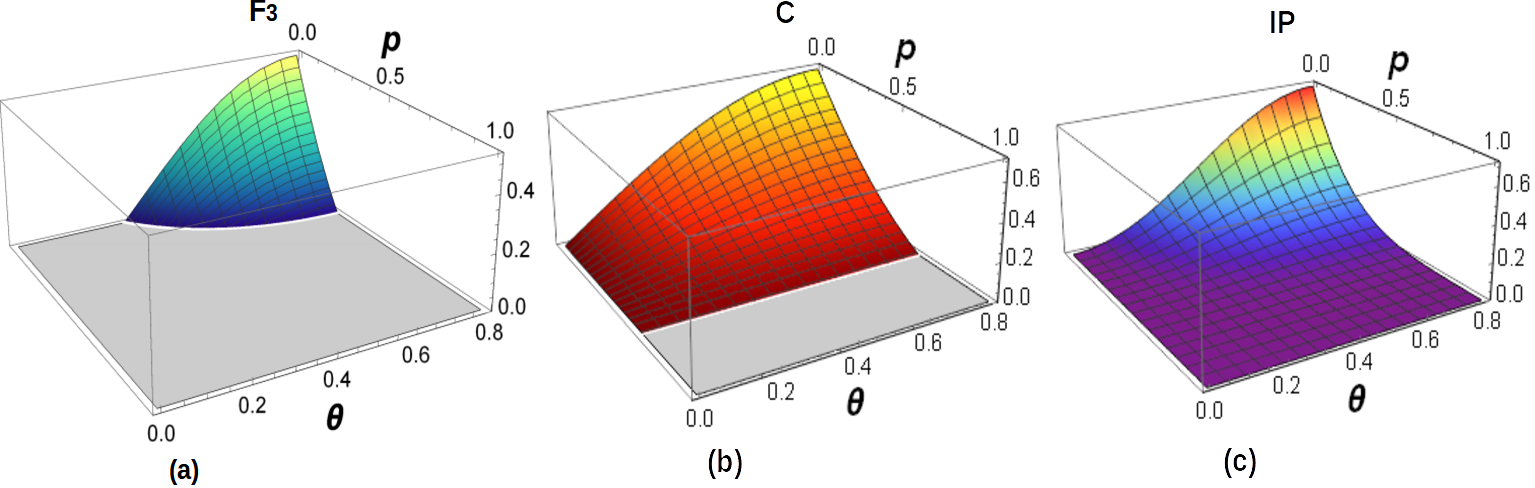}
\caption{(a) EPR-steering, (b) entanglement and  (c) interferometric power under SDC noise. fixed $k=4/5$.}
\label{fig:SDC_2}
\end{figure*}

As supplementary information to the results shown in the main text, we present another scenarios for the behaviour of the three analysed quantum properties. In \autoref{fig:PD_1} the effects of PD are shown for $k = 4/5$ and in terms of the noisy parameter and $\theta$. \autoref{fig:GAD_revival_EN} shows the death and revival phenomenon for entanglement whilst EPR-steering remains zero during the whole set of parameters. And, \autoref{fig:SDC_2} presents SDC effects for a fixed $k$ and in terms of $\theta$ and noisy parameter.


\end{document}